\documentclass[fleqn,12pt]{wlscirep}
\usepackage{epstopdf}
\usepackage{accents}

\usepackage{graphicx}
\usepackage{caption}
\usepackage{subcaption}

\newcommand{\fracc}[2]{\, \displaystyle \frac{ #1}{ #2}}
\newcommand{\all}[2]{\,\begin{align}
#1 
\label{#2}
\end{align}
}

\title{Effectiveness  of Alter Sampling in Social Networks}

\author[1,*]{Naghmeh Momeni}
\author[2]{Michael G. Rabbat}
\affil[1]{MIT Sloan School of Management}
\affil[2]{Department of Electrical and Computer Engineering, McGill University}

\affil[*]{nmomeni@mit.edu}

\begin{abstract}

Social networks play a key role in studying various individual and social behaviors. 
To use social networks in a study, their structural properties must be measured. 
 For offline social networks, the conventional procedure is   surveying/interviewing a set of randomly-selected  respondents. 
 In many practical  applications, inferring the network structure via sampling   is too prohibitively costly. 
 There are also applications in which it simply fails. 
For example, for optimal vaccination or employing  influential spreaders for public health interventions, we need to efficiently and quickly target well-connected individuals, which random sampling does not accomplish.
 In a   few studies, an alternative sampling scheme (which we dub `alter sampling') has proven useful.
  This method simply targets randomly-chosen neighbors of the randomly-selected respondents.
A natural question that arises is: to what extent does this method generalize?

Is the method suitable for every social network or only the very few ones considered so far? 
  In this paper, we  demonstrate  the  robustness of this method across a wide range of networks with diverse structural properties. 
The method outperforms random sampling by a large margin for a vast majority of cases. 
We then  propose an estimator to assess the advantage of choosing alter sampling over random sampling in practical scenarios, and  demonstrate its accuracy via Monte Carlo simulations on diverse synthetic networks. 
\end{abstract}
\begin{document}

\flushbottom
\maketitle
%
%
\thispagestyle{empty}









\section*{Introduction}

Social networks  are mathematical tools for modeling social relations and interactions, and for studying the interplay between structure and agency. 
They are employed in studying various social phenomena, such as contagion of health behaviors and  the 
adoption of new ideas and behaviors~\cite{coleman1957diffusion,banerjee2013diffusion,christakis2013social},  the spread of infectious disease~\cite{liljeros2003sexual,pastor2015epidemic},  the diffusion of information~\cite{yang2010modeling,lerman2010information},  and the  effect of network position and connections on individuals' power~\cite{coleman1988social,burt2000network}, job opportunities~\cite{granovetter1973strength},   cooperation~\cite{rand2011dynamic},  mental health~\cite{kawachi2001social}, longevity~\cite{olsen1991social}, 
 behavioral and ideological influence~\cite{kitsak2010identification,morone2015influence,aral2012identifying},  and migration decisions~\cite{massey1993theories,epstein2008herd}.  
 
 Descriptive studies of social networks relate the observed behavior of  a  social dynamical process or individual trait to the structural properties of the social networks.  Recent  studies   also   seek to leverage the theory of social networks  for practical applications, such as  `seeding' strategies and finding influential spreaders~\cite{morone2015influence},  public-health interventions~\cite{Lancet},   and for early detection of epidemic outbreaks~\cite{christakis2010social}. 
 This paper focuses on a specific practical method, which we call `alter sampling', that economically targets influential nodes while remaining agnostic of the network structure. We first briefly review a few successful applications. We then  provide a case study  using alter sampling on various social networks with different structural properties, and show that it performs remarkably well on all of them. Finally, we propose estimators for the advantage of using alter sampling over random sampling in reaching high-degree nodes. We conclude by discussing the implications of the effectiveness of alter sampling on how social networks are organized. 
 
  In social network studies,  descriptive or practical, analysis is carried out in terms of standard network statistics, i.e.,   quantities that pertain to the  structural properties of the social networks (e.g., degree, measures of centrality, clustering, homophily).  These properties  need to be observed and measured first. 
    Unlike some  networks with non-social origins (e.g.,  the Internet and the  World Wide Web),   measurements  in offline social networks are costly and challenging.  Efficient sampling and inference methods are needed to meet the  specific challenges  of social networks.

In practice, there  are situations where, due to time or budget constraints, or other practical concerns,  a  sampling  procedure would be unfeasible. 
As an illustrative example, 
consider the problem of vaccinating individuals against some disease in a village, where the vaccine resources are limited and we have to choose a small fraction of the population for immunization. 
It would be ideal to have complete knowledge over the network structure to identify the targets optimally. 
 Considerable time and resources would be needed for acquiring such complete knowledge of the network structure, which is not practical.
 So  we need to devise an efficient strategy to  identify the targets   without requiring knowledge of the social structure. The fewer questions the strategy requires  us to ask the respondents, the better.  
 
  A cost-effective, but naive vaccination strategy would be to randomly choose individuals for vaccination.  
  Intuitively, we need to find an efficient way to identify and vaccinate  the   well-connected individuals, because  if they get infected, they will transmit the disease to many people. 
   
  Random sampling does not necessarily capture these individuals because  it does not systematically target them. 
 It has been shown that a very effective strategy is  randomly selecting individuals, asking them to name someone they know, and then vaccinating those that are mentioned.
 This scheme is called \emph{acquaintance immunization}~\cite{cohen2003efficient,madar2004immunization,gallos2007improving}, and simulations show that despite  its  remarkably simple procedure,  it is  highly effective. 
   In this paper,  to use a  more broader term  that is also applicable in non-epidemics contexts,  we use the term \emph{alter sampling} to refer to the  method of random selection of  neighbors of a random sample.

   The promising feature of  alter sampling  is that it targets influential nodes with almost no knowledge of the structure of the underlying social structure.  
Christakis and Fowler~\cite{christakis2010social}  describe  an empirical study using the idea of alter sampling to monitor the spread of the H1N1 flu.

Comparing two samples of students, one obtained via  random  sampling and one via  alter sampling, 
 they showed that the  prevalence curve 
 for the  latter  sample  is shifted 13.9 days forward in time as compared to the former.  This indicates that alter sampling can be utilized for the  detection of outbreaks in the early stages of an epidemic.

The idea of using alter sampling for the early detection of outbreaks can be extended beyond infectious diseases, and can also be applied to information contagion. 
The   diffusion of viral online content on Twitter is  an    example, where it is shown that 
samples of users obtained by alter sampling (refered to as \emph{sensors}) receive   viral hashtags  earlier  than samples obtained by random sampling. 
The difference still remains after controlling for possible reverse causality 
 (that sharing viral content is not the result, but the cause of network position) by showing that virality of posts and network position of the post originator are not significantly correlated~\cite{garcia2014using}.

Alter sampling can also be  used to promote the spread of information and awareness in social networks.  In a   very recent study, this method was used for improving the impact of health intervention in 32 villages in Honduras~\cite{Lancet}. Two distinct health interventions (one   nutritional and the other pertaining to water purification) were made. 
The products and instructions were given to 5\% of the population
(reached via random sampling in some villages and via alter sampling in others). 
The final prevalence of the adoption  and the  general   knowledge  of  the  health behaviors were greater in villages where alter sampling was used (about 12\%).  

The above empirical observations  suggest that alter sampling is a potent and efficient practical method for finding influential nodes. 
To be able to confidently use it in practice, we need to verify that the success of the method was not due to peculiarities of the above (very few) cases, and to 
  ascertain  its robustness across a wide range of networks with social origin. 
 This is the first focal task of the present paper. We demonstrate that alter  sampling  is robust  in a range of networks with social origins
 with diverse structural properties  (we  consider positively, negatively, and neutrally assortative networks, high and low degree variance, different levels of clustering and density). We demonstrate that  alter sampling performs well across all of them, and performs remarkably similarly.  
 This sheds light on micro mechanisms that are present in networks  with social origin that do not depend on the specific properties of the context.
 
 When employing alter sampling, one may also wish to estimate how much better of a sample has been obtained by using alter sampling instead of uniform random sampling.
 That is, we need to quantify  the benefit of  using  alter sampling as compared to random sampling.  
 Since  we consider scenarios in which the structure of the social network is unknown, we cannot use any  structural information
 to estimate the benefit of using alter sampling, either a-priori or retrospectively. 
 For example, after using interview data to vaccinate the alters that respondents nominate, how can we assess the gain of this method over uniform random sampling?
 Answering this question is the second focal task of the present paper.  
 We propose estimators that use  interview  data to quantify the `gain' of alter sampling which is the advantage of this method over random sampling in identifying higher-degree nodes.

\section*{Results}

We can quantify the performance of alter sampling in various ways.
The most basic individual attribute that characterizes the  influence of a node on  a dynamical process running over the network is the degree. Thus  we take the  expected value of the degrees of the nodes reached via a sampling scheme as its merit, and the gain of choosing one sampling scheme over another is quantified by the relative increase in this merit.  
 With the few recent exceptions of employing alter sampling in practical settings, most studies have employed random sampling. In the present paper, we seek to quantify how much loss is associated with such a decision. 
For undirected networks, we compare the degree of  node $x$ with the mean degree of its neighbors. The ratio of these two quantities is the local   gain of     choosing alter sampling over random sampling.  For node $x$, we denote this ratio by $\mathcal{G}_x$. The average value of this ratio over all nodes, which we denote by $\mathcal{G}$,  gives the expected gain. Choosing alter sampling is justified if $\mathcal{G}>1$.  
In directed networks, if node $y$ follows node $x$ (that is, there is a link from node $y$ to node $x$), then $y$ is called the in-neighbor of $x$, and $x$ is called an out-neighbor of node $y$. 
The number of in-neighbors and out-neighbors  of a node are called its in-degree and out-degree, respectively. 
For directed networks, we only consider in-degrees, because social influence operates in the direction of links. 
That is, for a given node $x$, it is the in-neighbors of $x$ that are influenced by $x$, and not the out-neighbors. Otherwise, the methodology is similar to the case of undirected networks.

The descriptions of the data sets used are provided in the Methods section. Their summary statistics are presented in Table~\ref{tab:dir} (for directed networks) and Table~\ref{tab:undir} (for undirected networks). 

\begin{figure}[!h]
        \centering
                \includegraphics[width=.9 \columnwidth]{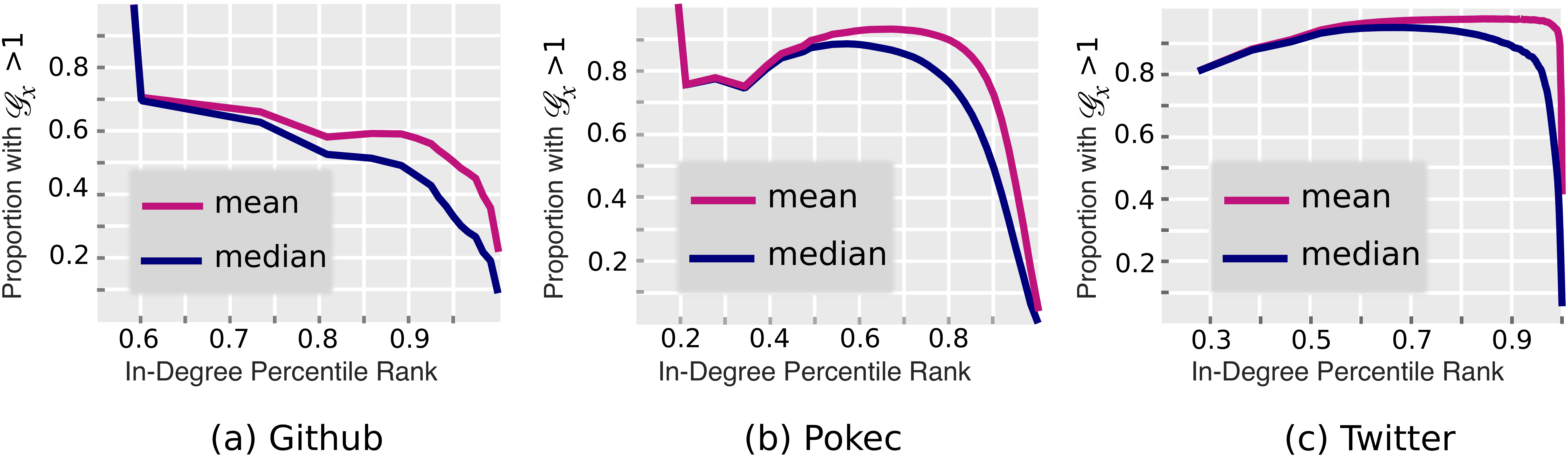}
                \caption{Empirical distribution of nodes with $\mathcal{G}_x>1$ as a function of in-degree percentile rank for directed networks. }
                \label{fig:dir}
\end{figure}

\begin{figure}[!h]
        \centering
                \includegraphics[width=\columnwidth]{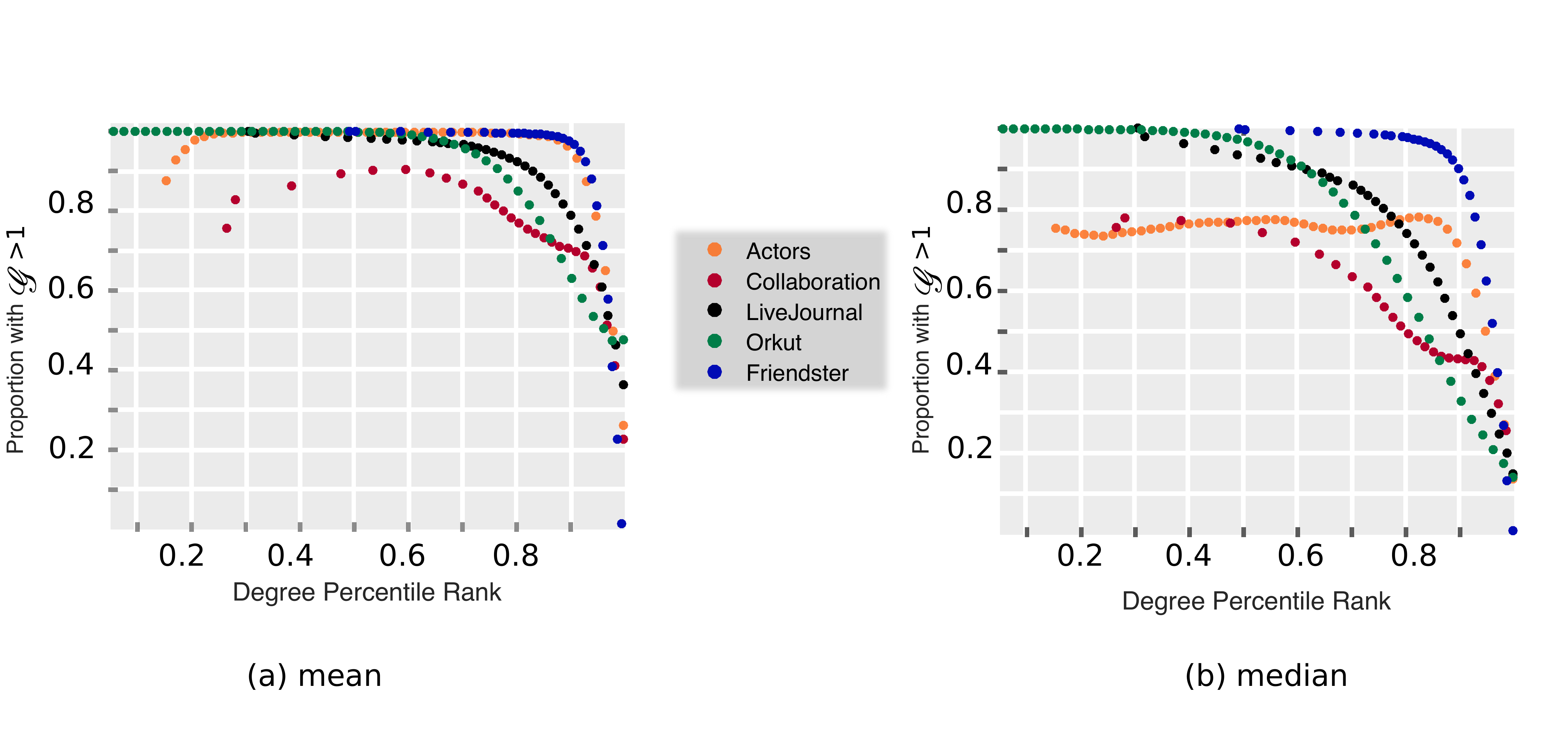}
        \caption{ Empirical distribution of nodes with $\mathcal{G}>1$ as a function of degree percentile rank for undirected networks. In (a), gain is defined using the mean, and in (b), using the median, as discussed in the text. }\label{fig:undir}
\end{figure}

There are various ways we can quantify the advantage of using alter sampling instead of random sampling. In Figure~\ref{fig:dir} and Figure~\ref{fig:undir}, we depict the proportion of nodes in each degree-percentile for whom $\mathcal{G}_x>1$  in  directed and undirected networks, respectively. It can be observed that for all networks, a vast majority of nodes do meet this criterion. For all directed networks under consideration, over 85\% of the population have $\mathcal{G}_x>1$. For undirected networks, this number is even higher (near 95\%). Note that this phenomenon  holds  whether we   compare the degree of each node with the mean or with the median degree of the neighbors, which  highlights the robustness of the observed phenomenon against possible outliers (hubs).

\begin{figure}[!h]
        \centering
        \begin{subfigure}[b]{0.47 \columnwidth}
                \includegraphics[width=\columnwidth]{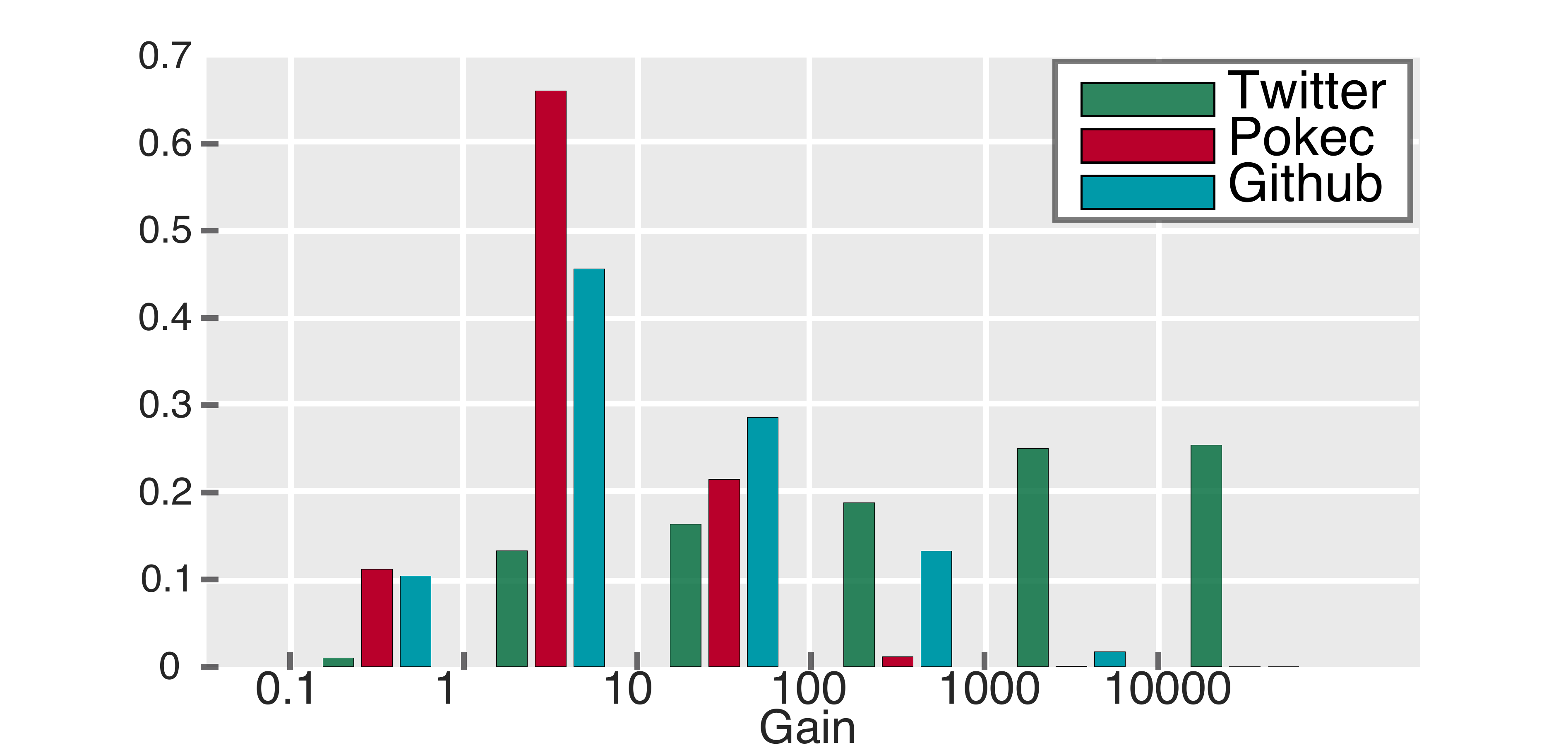}
                \caption{
                directed networks (mean)}
                \label{histgaindir}
        \end{subfigure}%
        ~ 
        \begin{subfigure}[b]{0.47  \columnwidth}
                \includegraphics[width=\columnwidth]{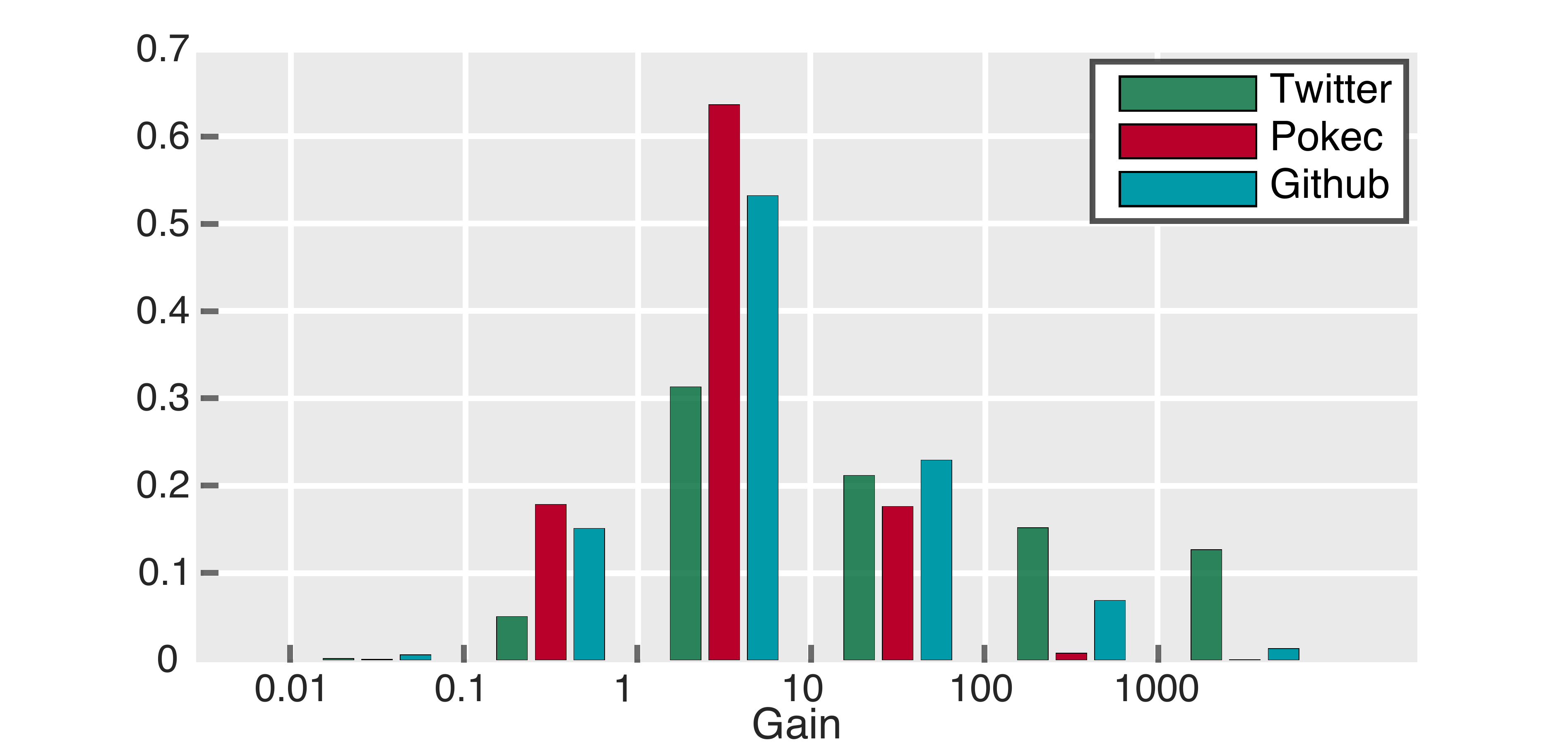}
                \caption{
                directed networks (median)}
                \label{histdirgainmed}
        \end{subfigure}%
        \\
     ~   \begin{subfigure}[b]{0.47 \columnwidth}
                \includegraphics[width=\columnwidth]{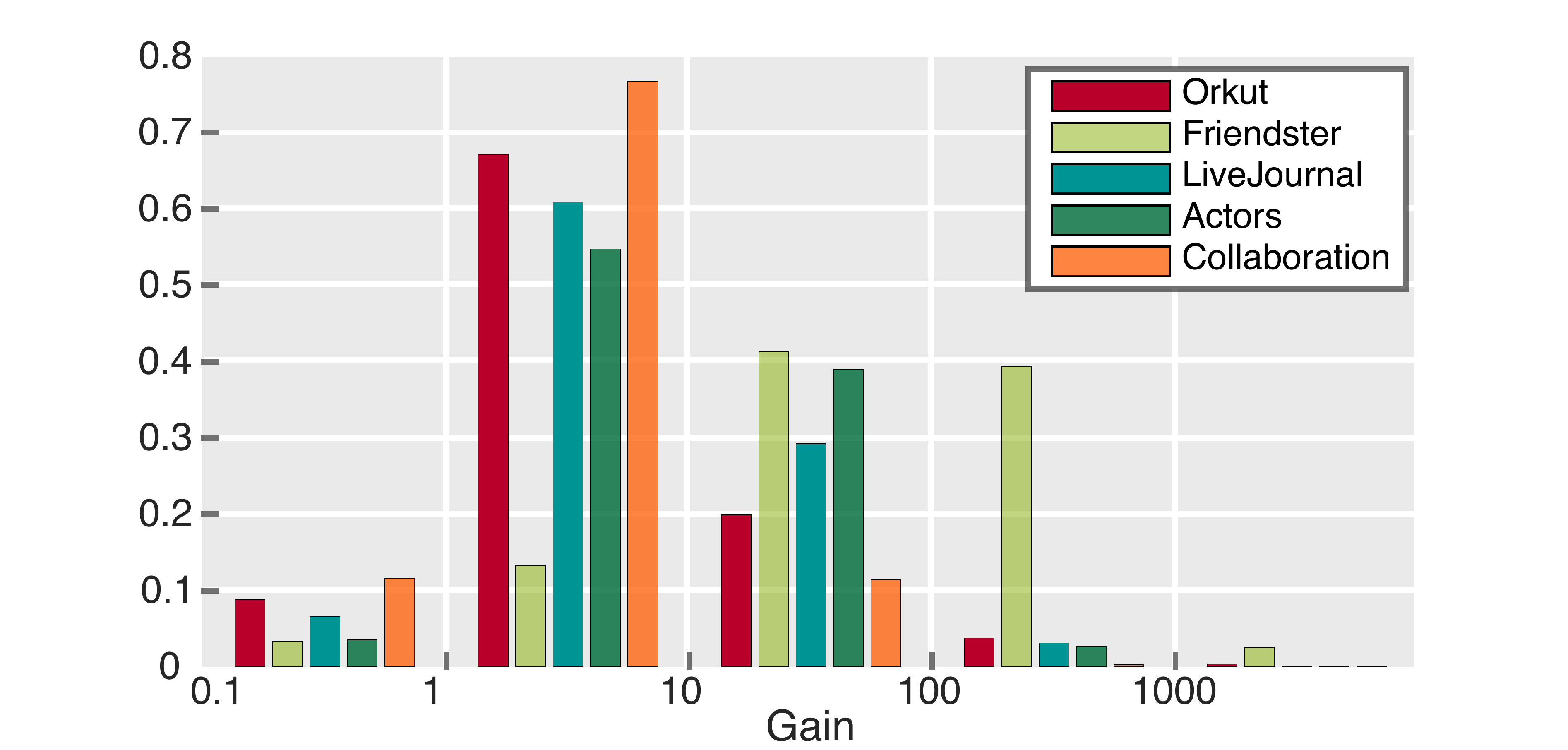}
                \caption{
                undirected networks (mean)}
                \label{histgainundir}
        \end{subfigure}%
        ~ 
        \begin{subfigure}[b]{0.47 \columnwidth}
                \includegraphics[width=\columnwidth]{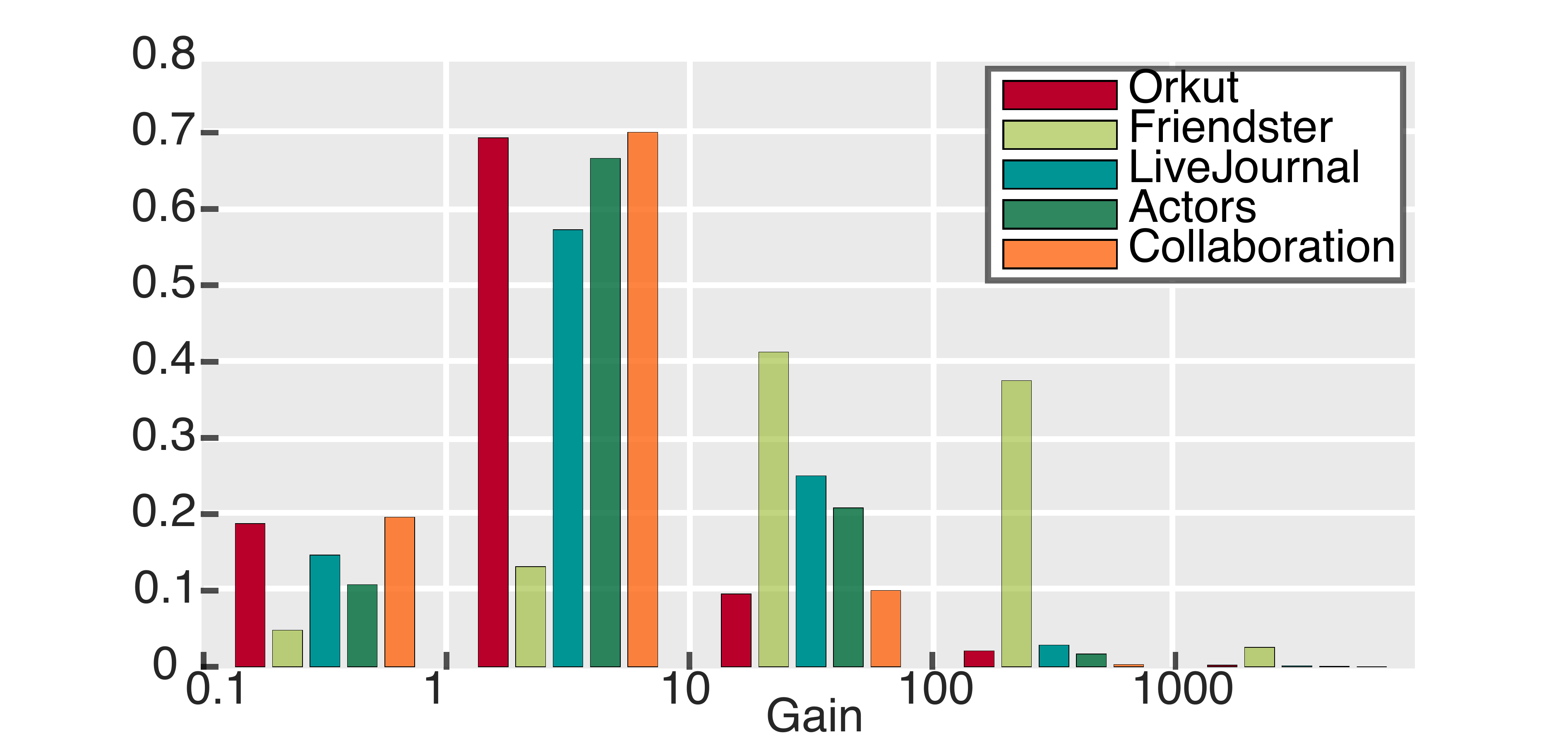}
                \caption{
                undirected networks (median)}
                \label{histundirgainmed}
        \end{subfigure}
        \caption{\footnotesize{The distribution of $\mathcal{G}_x$ for different networks. The top row pertains to the directed  networks and the bottom row pertains to the undirected networks. In the left column, the gain for each node is defined as the average degree of its neighbors to its own degree. In the right column, the median degree of the neighbors is used instead of the mean. } }\label{G1}
\end{figure}

So far we demonstrated  that the observed seemingly-universal advantage of using alter sampling in social networks is not attributable merely to hubs. 
The above  measures  for  prevalence of $\mathcal{G}_x>1$  indicate that  for a high proportion of the population,  alter sampling is superior to  random sampling, but these measures do not quantify to what extent that is so. To investigate this, we plotted the histogram of $\mathcal{G}_x$ across all nodes for different networks in Figure~\ref{G1}. It can be seen that across all these networks, the distribution of $\mathcal{G}$ is highly skewed, that is, there exist nodes for which the gain of using alter sampling is overwhelmingly large, and for the  majority of nodes this gain is still considerably large (that is, $O(10^1)$ gain for alter sampling). As before, the gain is robust against outliers, since using the median instead of the mean to define the gain does not change the results significantly. 

\begin{figure}[!h]
        \centering
                \includegraphics[width=\columnwidth]{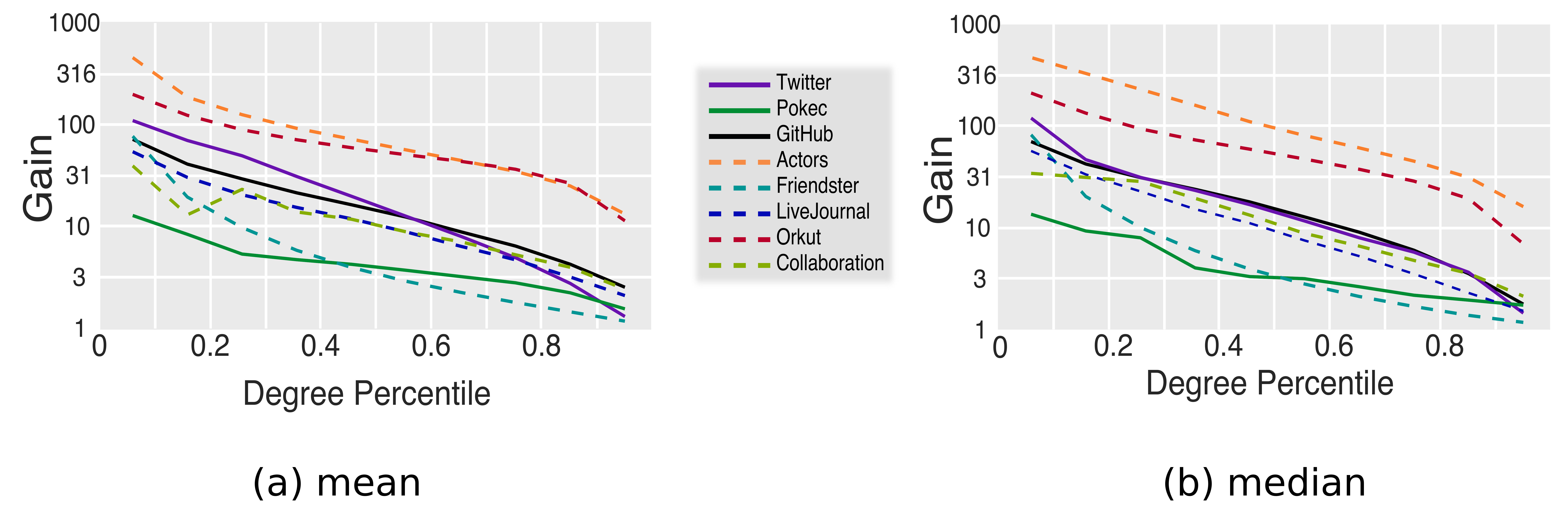}
                \label{gainperc}
        \caption{ The gain of performing alter sampling as a function of the degree percentile of the target node. In (a), the average degree of neighbors is compared to the degree of each node to define the gain, and in (b), the median is used.   }\label{G2}
\end{figure}

 Furthermore, Figure~\ref{G2} shows that the average  $\mathcal{G}_x$  value  as a function of their degree percentile. 
   It can be seen that in all networks, the  gain
  is considerably high for a vast majority of nodes. 
  Note that the behavior of the gain  function is similar across all networks, whereas their structural properties  (such as  assortative mixing, clustering, density, average degree, and degree variance) 
  are widely different, as reported in Table~\ref{tab:dir} and Table~\ref{tab:undir}. 
  This suggests  that alter sampling is  considerably robust to variation of network structure. This endows alter sampling with a notable versatility,  thus it can be reliably used in practice for cases where it is not feasible to obtain the structure of the underlying social network through standard methods of social network studies, such as interviewing and surveying the population.

\begin{table}[!h]
\centering
\begin{tabular}{cccccccc}
 &  & GitHub &  & Pokec &  & Twitter &  \\ \hline \hline
$N$ &  & 46423 &  & 531478 &  &  5489933  &  \\
$E$ &  & 156280 &   &30622564  &  & 193245641 &  \\ 
$\overline{k}$ &  & 3.366 &    &18.754  &  & 35.2 &  \\
$\hat{k}_{in}$ &  & 1 &  & 8 &  & 4 &  \\
$\sigma_{in}$ &  & 20.25 &    &32.140  &  & 989.01 &  \\
 \hline  
\end{tabular}
\caption{
Directed networks: $N$ and $E$ are number of nodes and number of edges, respectively. $\overline{k} $, $\hat{k}_{in}$ and $ \sigma_{in}$ denote average in-degree (which is equal to average out-degree), median of in-degrees and standard deviation of in-degrees, respectively. }
\label{tab:dir}
\end{table}

\begin{table}[!h]
\centering
\begin{tabular}{cccccc}
                  & Actors & Collaboration & LiveJournal & Friendster &Orkut  \\ \hline\hline
$N$    &  894615     & 69032      & 3997962    &  22493449    & 3072441      \\
$E$  & 57060378    & 450622  & 34681189   & 180606713   &  11785083 \\
$\overline{k}$         &  127.5   &  13.05     & 17.40  &   16.058  &    76.28  \\
$\widehat{k}$     &     41   &    5      &   6   &    3     &    45    \\
$\sigma_{k}$      &  317.5 &  27.97   &  42.95   & 53.29  & 154.78 \\
$r_{kk'}$         &  0.20  & 0.6018  &  0.045  & -0.1816  &0.0158 \\
$\overline{C}$        &  0.4724   & 0.5977  &   0.2842 &  0.0734 & 0.1666\\ \hline 
\end{tabular}   
\caption{Undirected networks: $N$ and $E$ are number of nodes and number of edges, respectively. $\overline{k} $, $\hat{k}$ and $ \sigma_{k}$ denote average degree, median of degrees and standard deviation of  degrees, respectively. Degree assortativity and average clustering coefficient of the graph are denoted by $r_{kk'}$ and $\overline{C}$, respectively.  }
\label{tab:undir}
\end{table}

Now we attend to an important practical question, that is, to estimate the gain of  choosing alter  sampling  over random sampling from empirical data. 
We provide  an estimator for the gain in choosing alter sampling over random sampling. In  practice, an offline social network study (such as those that would be needed for optimal vaccination) involves interviewing people and asking them to nominate alters.   Suppose that we also ask people to report the number of people they know~\cite{hill2003social}. We aim to estimate the gain of alter sampling from the sequence of degrees that the respondents provide. Note that we cannot ask a respondent about the degree of their alters, because although people generally have a good knowledge of their own social ties, they might not be necessarily good at  providing reliable  estimates for  the number of ties  of one   of their friends.

Suppose that the underlying network has degree distribution $p(k)$.  This means that with probability  $p(k)$, a randomly-chosen respondent  has degree $k$. 
 Suppose that a node with degree $\ell$ is mentioned as an alter. There are $Np(\ell)$ nodes of degree $\ell$ in the network, where $N$ is the network size. Each of these nodes has $\ell$ neighbors that could be the initial respondent. Thus, there are on average $N\ell p(\ell)$ nodes that could have mentioned a degree-$\ell$ node.  Denoting the mean degree of the network by $\mu_1$,  the probability that a mentioned alter has degree $\ell$ is given by $ N \ell p(\ell) /\sum_\ell N \ell p(\ell)= \ell p(\ell) /{\mu_1}$. 
 We use these conditional probabilities to obtain the expected degree of a named 
 alter, which is  $  \sum_\ell \ell^2 p(\ell)/\mu_1= \mu_2 / \mu_1$, where $\mu_2$ is the second moment of the degree distribution. For node $x$, the gain of alter sampling is simply the ratio of the expected alter degree to that of node $x$. Thus the expected gain is given by $ (1/N) \sum_x  [(\mu_2/\mu_1)(  1/{k_x})]$. Let us denote the harmonic mean of the degrees by $\mu_{h}$. That is, we have $\mu_{h}=\sum_k k^{-1} p(k)$.  Thus, the expected gain is given by $\mathcal{G}= \mu_2 \mu_h /\mu_1$.  Denoting the set  of  respondents by $\mathcal{R}$, the total number of respondents by $r$,  and the reported degree of respondent $i$ by $\tilde{k}_i$,  we arrive at the following estimator for the gain of choosing alter sampling:
 \all{
 \widehat{\mathcal{G}}=  \fracc{ \left(  \displaystyle \sum_{i \in \mathcal{R}} \tilde{k}_i^2 \right) \left( \displaystyle \sum_{i \in \mathcal{R}} \frac{1}{\tilde{k}_i} \right) }{\displaystyle r \sum_{i \in \mathcal{R}} \tilde{k}_i}
 .}{e1}

\begin{figure}[!h]
        \centering
        \begin{subfigure}[b]{0.45 \columnwidth}
                \includegraphics[width=\columnwidth]{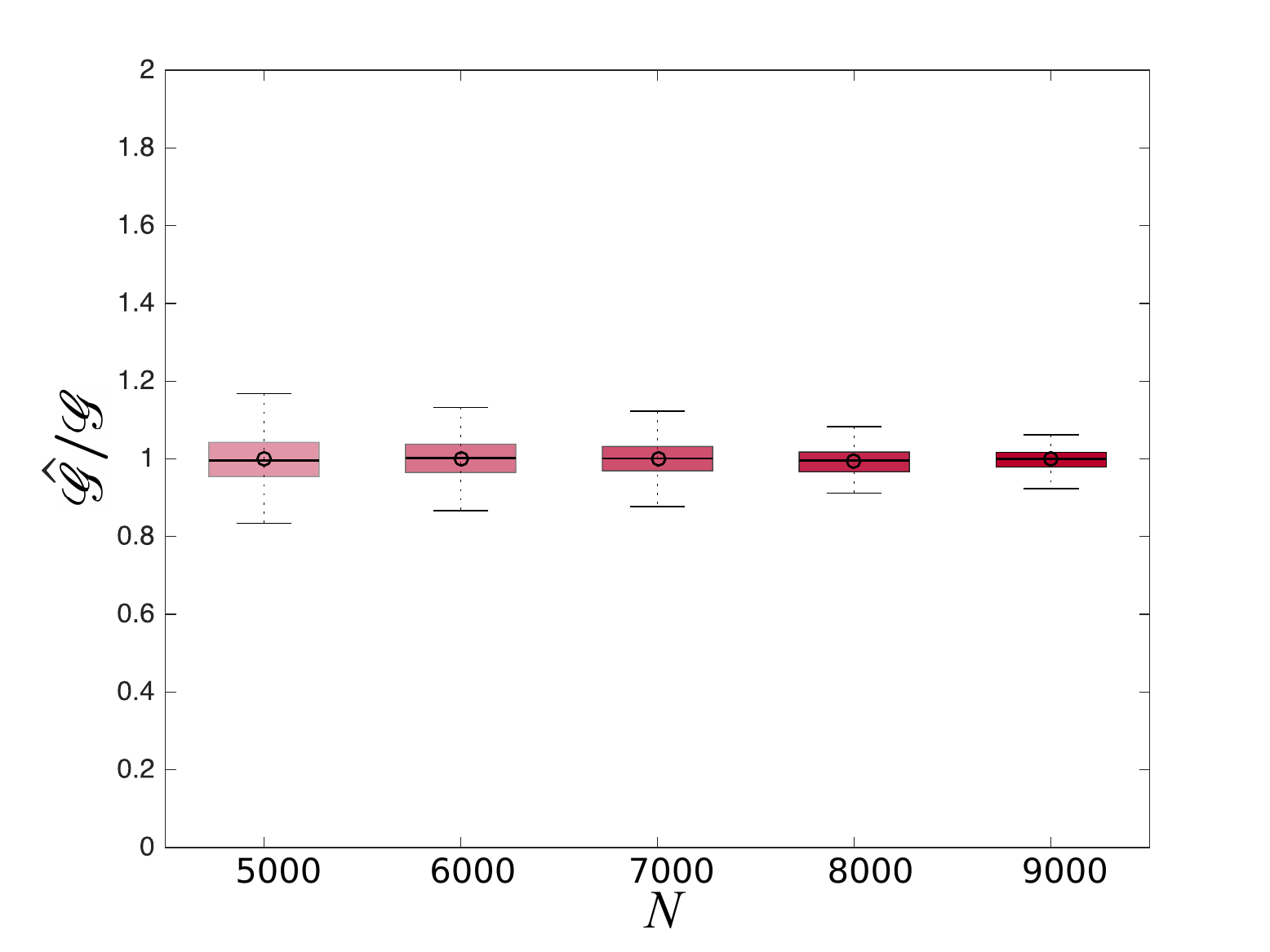}
                \caption{
                small-world model (high clustering, low average path length)}
                \label{SW1}
        \end{subfigure}%
        ~ 
            \begin{subfigure}[b]{0.45 \columnwidth}
                \includegraphics[width=\columnwidth]{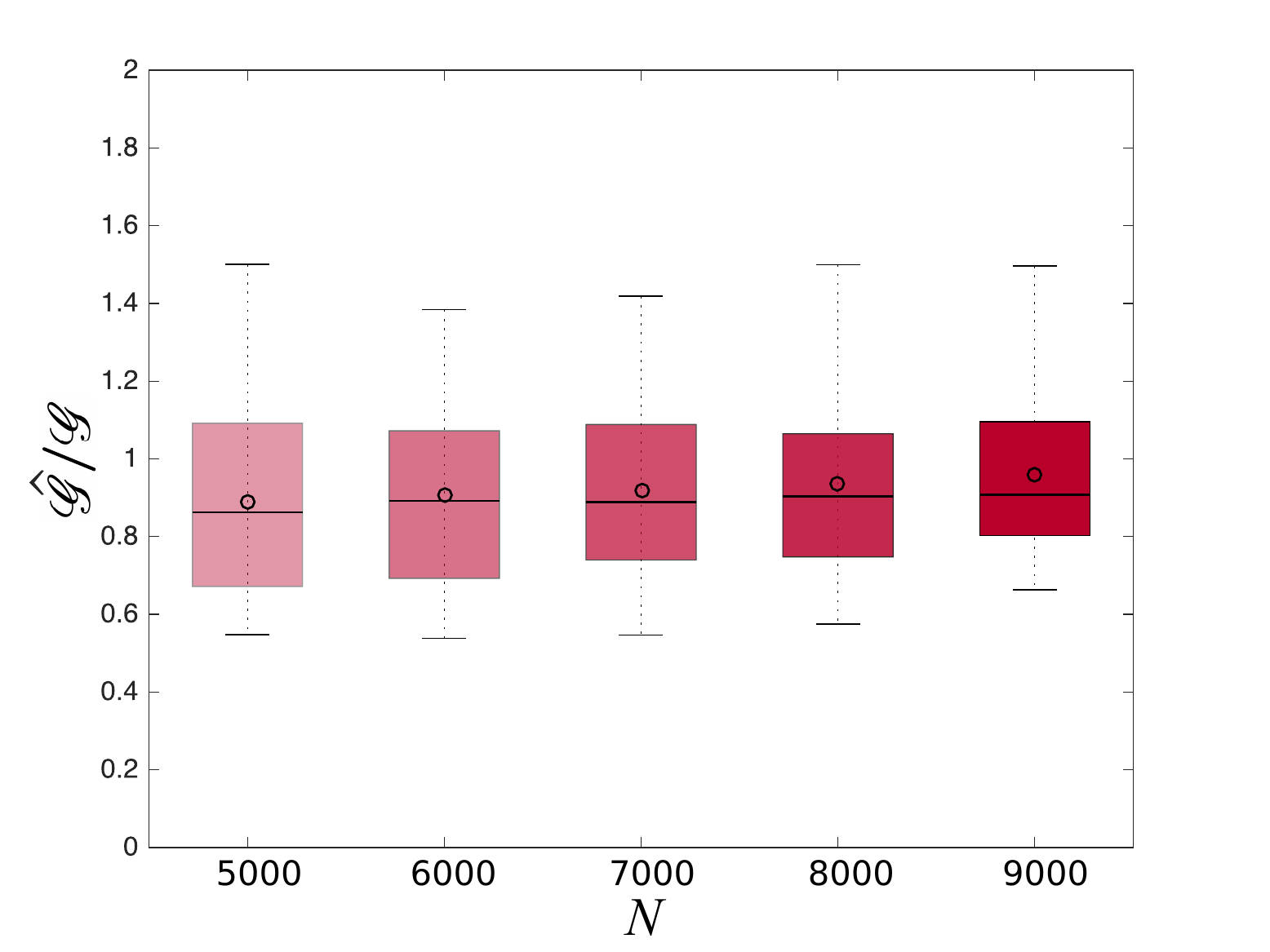}
                \caption{
                preferential attachment model (low clustering, heavy-tailed degree distribution)}
                \label{BA1}
        \end{subfigure}   
        \\
         \begin{subfigure}[b]{0.45  \columnwidth}
                \includegraphics[width=\columnwidth]{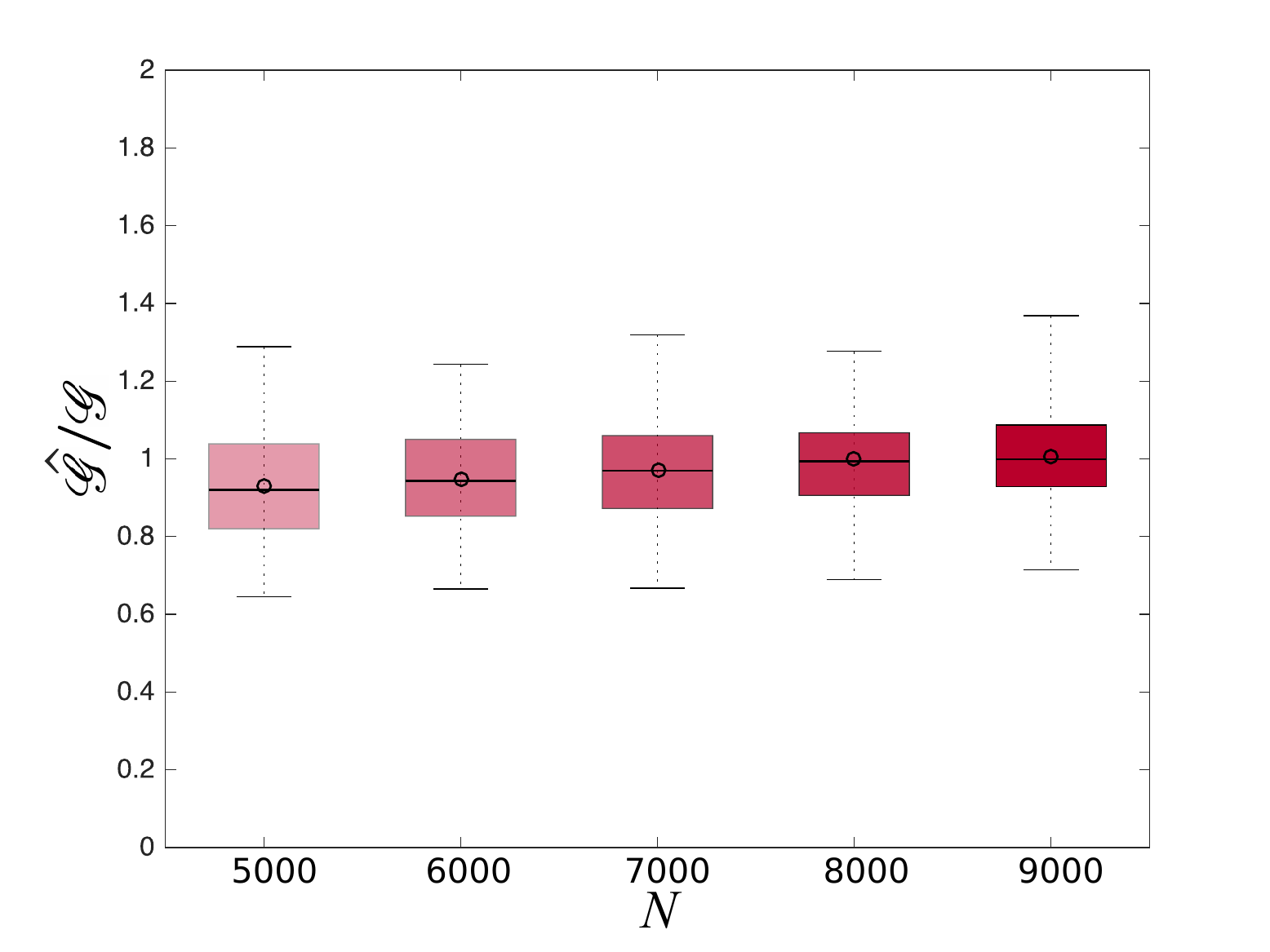}
                \caption{
                HK model (high clustering, heavy-tailed degree distribution)}
                \label{HK1}
        \end{subfigure}%
 ~   \begin{subfigure}[b]{0.45 \columnwidth}
                \includegraphics[width=\columnwidth]{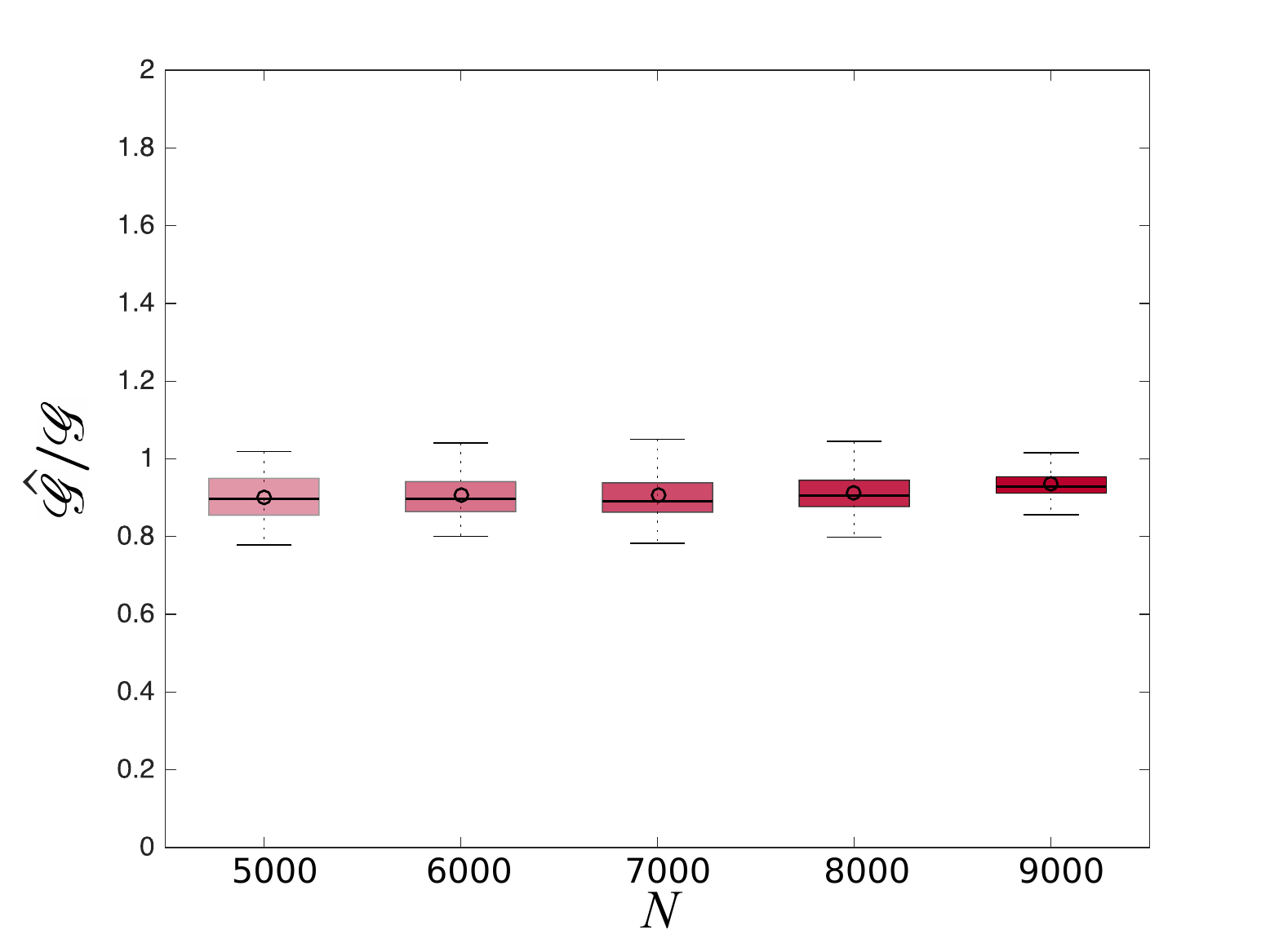}
                \caption{
                KE model (high clustering, low average path length, heavy-tailed degree distribution)}
                \label{KE1}
        \end{subfigure}%
        ~ 

        \caption{ The performance of the first proposed estimator for the gain of alter sampling for different families of networks.   }\label{BP1}
\end{figure}

To assess the  performance  of the estimator, we would ideally need an empirical sampled networked data set via the above mechanism for which the real underlying network is also known. Every existing offline social network data set in the literature is already the sampled version, and the true underlying network of interactions is not known for any of them. Noting that this caveat prevents testing the estimators on real networks, we use synthetic networks. 
We use network models from the literature that have been proposed to emulate properties that have been observed in social networks. One such model is the small-world network model~\cite{watts1998collective}. This 
model was proposed in order to  capture  two important structural features observed widely in real social networks: high clustering, and small average path length. The  former  captures the high transitivity that is typical in networks of social origin (that is, friends of a person tend to become friends with high probability), and the latter pertains to the well-known six-degrees-of-separation phenomenon (that every two persons in society are connected via a   very  short chain of acquaintances). 
  We synthesized 10000 networks (see Methods for details of the generation process of all network models considered) and for each case  we estimated the gain from the above estimators and calculated its ratio to the true value. The closer this ratio is to unity, the better the estimators are performing. Figure~\ref{SW1}    presents the results.  It can be observed that the estimators are performing with acceptable accuracy, with errors mostly less than 10\%. 

The second conventional network generation model  is the preferential attachment model. 
 Proposed in~\cite{price1976general} and later in~\cite{barabasi1999emergence},  this model  emulates the empirically-observed  heavy-tailed nature of the degree distributions in  diverse  networks  (such as the network of scientific citations, scientific collaborations, and the worldwide web).   The results for this model are presented in Figure~\ref{BA1}.

  The third generative network model    that we use is the one proposed by Holme and Kim~\cite{holme2002growing}. The model  combines the   preferential attachment model of network growth with high clustering. We refer to this model as the HK model. The HK model  adds a triad-formation step to  the conventional preferential attachment model, and makes it more suitable to modeling networks of social origin than the basic preferential attachment model (which has vanishing clustering coefficient for large networks).  The results for  the HK  model  are presented in Figure~\ref{HK1}. It can be observed that the variance of the estimator is slightly higher than it was for small-world networks, but it is slowly decreasing with network size. 

The  fourth  model, which we refer to as the KE model ~\cite{klemm2002growing}, yields small average path length in addition to high clustering and skewed degree distribution.  Similar to the previous models, the estimator has an error of less than 10\% in the majority of the simulation trials. The results for the KE model are presented in Figure~\ref{KE1}. 

In all the simulation trials, the fraction of randomly-chosen respondents (whose random neighbors then constitute the alter set) are chosen uniformly at random between 0.1 and 0.2. So, equivalently,  the value of $r$ in Equation~\eqref{e1} is randomly selected between 10\% and 20\% of the total population.


%
%
%
%

\section*{Discussion}

The presented results suggest that alter sampling is a strong and economical method for targeting well-connected nodes in the network when standard sampling procedures are   costly and infeasible.   The results demonstrate a remarkable versatility and robustness of this method. 
We considered many network data sets with large size and diverse structural properties (positive,  negative, and neutral assortative mixing, high and low clustering and density, high and low variance of degrees and average degree), and in all cases, alter sampling is advantageous over  random  sampling for a vast majority of nodes. This holds even if we consider the median of the degrees of neighbors instead of the mean in order to define the gain of choosing alter sampling over random sampling. Hence, although in the literature this phenomenon has been linked to the presence of hubs, our results indicate that a more prevalent structural feature must exist in all these networks to give rise to this behavior.

We also proposed an estimator to assess the gain of choosing  alter sampling over random sampling in practical scenarios and investigated its performance on synthetic networks generated via four distinct conventional network generation models.  We observed that the proposed estimator performs remarkably well across a diverse range of structural parameters of the synthetic networks. 

The immediate extensions to more practical scenarios would be to consider imperfect response (due to, for example, forgetting or fatigue). Also, in some cases it might not be feasible to ask respondents to count the number  of their friends. It is time-consuming and there might be situations in which there is only time to ask a few alter names. In this case, we will have to estimate $\mu_2$, $\mu_1$, and $\mu_h$ from the response data. Usually, there is a cutoff on the number of alters each respondent must mention, which is typically less than 10. In this case, the above moments of the degree distribution must be estimated from a dataset in which for each node only about 10\% of the links are known. This is an interesting problem of statistical inference with immediate practical importance. We hope the results presented in this paper will invite   closer investigations    of   alter sampling and its robustness and limitations, as well as the associated network sampling problems   that will be  practically imperative.

\section*{Methods}
\footnotesize

\subsection*{Network Models}
\textbf{Small-world:} We use  a  variant~\cite{newman1999renormalization}  in which the network is built as follows: we  begin with a $2b$-regular lattice (a ring in which each node is connected to $b$ immediate neighbors from each side), and we create  each non-existing link  with constant probability $p$, independently.  Since it has been  consistently shown in the literature that cognitive constraints limit the effective number of social ties a human can actively maintain to about 150~\cite{hill2003social,zhou2005discrete,dunbar2015structure,fuchs2014fractal} (also called the \emph{Dunbar number}), we 
restrict the space of parameters to a domain for which the average degree is about 150. The value of $b$ was randomly chosen between 5 and 10, and the value of $p$ was chosen in a way to yield the average degree no greater than 200.  \\
\textbf{Preferential Attachment:} In this model, nodes are added to the network sequentially, and each incoming node attaches to $m$ existing nodes that are selected with degree-proportional probabilities.  
 We selected $m$ randomly between 50 and 75, generating networks with average degree  between 100 and 150. We considered sizes from 5000 to 9000.
 We synthesized 1000 networks for each size.
 \\
 \textbf{HK:} The parameters of the model are the initial number of links that each incoming node creates when it is being added to the network, and the triad formation probability. In the ensemble of networks that we generated, we randomized the first parameter between 50 and 100, and the triad formation probability was randomly generated in the interval $[0,0.5]$, and a network was only accepted if the mean degree was less than 150. For each network size, we generated 1000 synthetic networks and implemented the sampling procedure described above with $G$ randomly chosen between 5 and 10, because values of $G$ more than 10 are rare in real social network studies~\cite{momeni2016inferring}. 
 \\
 \textbf{KE:} In this model, at each timestep these are $m$ active nodes and as a new node is added, it creates $m$ links. Each link, with probability $\mu$ connects to a random node chosen with degree-proportional probabilities according to the basic preferential attachment scheme, and with probability $1-\mu$ attaches to one  of the active nodes. The new node becomes active and one of the previously-active nodes becomes inactive with probabilities inversely proportional to degrees. This procedure is then repeated. We have randomized the parameter space with the restriction that the generated networks have mean degree between 100 and 200.

\subsection*{Data}

To ascertain the versatility of alter sampling, we considered five undirected and three directed networked data sets. A quantitative summary of their properties is presented in Table~\eqref{tab:dir} and Table~\ref{tab:undir},  for directed and undirected networks, respectively.  Below we provide a qualitative description of the data sets:
\\
 \textbf{  Film Actor Network:}
  We use a network derived from the IMDB movie/actor network available in the University of Florida Sparse Matrix Collection~\cite{davis2011university}. This bipartite network consists of 428,440 movies and 896,308 actors and stores the movies in which each actor has appeared. Based on this graph, we can build the co-starring network. In this network each node represents an actor and an edge connecting two nodes indicates that those nodes have co-appeared in at least one movie. Note that we do not consider weights for the edges.
  \\
 \textbf{Scientific Collaboration Networks}
 We use the collaboration network available at~\cite{leskovec2007graph}. The dataset is extracted from the e-print arXiv and covers scientific collaborations between authors papers in five categories in the period from January 1993 to April 2003. If an author $x$ co-authored a paper with author $y$, the graph contains a undirected edge from $x$ to $y$. 
\\
 \textbf{LiveJournal}
 LiveJournal is a social networking service where users can keep a blog, journal or diary and also can declare friendship with each other. The network that we use here is available at~\cite{backstrom2006group,leskovec2009community} and consists of about four million users.
 \\ 
 \textbf{Friendster}
Friendster is an on-line gaming network. Before re-launching as a game website, it was a social networking site. The network that we use in this paper is a subset of the graph available at~\cite{yang2015defining} consisting of more than 22 million nodes.
\\
 \textbf{Orkut}
 Orkut is a free on-line social network. The network used in this paper is available at~\cite{yang2015defining} and consists of more than three million users.
\\
  \textbf{Twitter}  For the network of Twitter users, we use the dataset  collected by Kwak et al. \cite{kwak2010twitter}. This dataset describes the connectivity among users who joined Twitter prior to August 2009. The subgraph that we use has 5.8 million users and more than 193 million edges.  
\\ 
 \textbf{GitHub}
 The site github.com offers free code repository hosting for public projects and paid code repository hosting for private projects. Individuals can follow one another, like users of Twitter, in order to stay aware of each other's activities. In~\cite{gium} the GitHub Archive
site\footnote{http://www.githubarchive.org/} was used to download past compressed archives of hourly activities over a one-year period. The collected and processed data are used to create multiple graphs including the followership graph which is used in this paper. In this graph  there is an edge from node $x$ to node $y$, if user $x$ follows user $y$.
\\
 \textbf{Pokec}  Pokec is the most popular on-line social network in Slovakia. The dataset is available at~\cite{takac2012data} and consists of more than 1.6 million nodes and more than 30 million edges.



\section*{Acknowledgements}
This work was supported, in part, by the Natural Sciences and Engineering Research Council of Canada (NSERC) grant RGPIN/341596-2012.

\section*{Author contributions statement}
 
 N.M. and M.R. conceived the research problem. 
 N.M. gathered  the data and performed the analyses. 
 N.M. and M.R. discussed the results, and wrote and  reviewed the manuscript. 

\section*{Competing financial interests}
The authors declare no competing financial interests.


%

\end{document}